\documentclass[a4paper,11pt]{article}
\PassOptionsToPackage{hyphens}{url}
\pdfoutput=1 

\usepackage{jinstpub}
\usepackage{subfigure}
\usepackage{lineno}
\title{A Portable Cosmic Ray Detector for School Education}

\author[a,b,1]{L. Bomben\note{Corresponding author.}}
\author[a,b]{, S. Capelli}
\author[a]{, C. Fanzini}
\author[a,b]{, E. Lutsenko}
\author[a,b]{, V. Mascagna}
\author[a]{, C. Petroselli}
\author[a,b]{, M. Prest}
\author[a,b]{, F. Ronchetti}
\author[a,b]{, A. Selmi}
\author[b]{and E. Vallazza.}

\affiliation[a]{Università degli Studi dell'Insubria - Dipartimento di Scienza e Alta Tecnologia, via Valleggio 11, Como, Italy}
\affiliation[b]{INFN - Sezione di Milano Bicocca, piazza della Scienza 3, Milano, Italy}

\emailAdd{lbomben@studenti.uninsubria.it}

\abstract{
This article describes the design, assembly and characterization of a portable cosmic ray detector, developed by the INSULAB group and suitable for teaching activities aimed at high school students. It consists of a compact aluminum suitcase containing three plastic scintillator modules coupled to photomultipliers, readout by a custom compact electronics chain and powered by a power bank. The modules operate in coincidence and the system records the arrival time of each particle and the time over threshold of the signal of each scintillator module. The data are acquired and processed by a Raspberry PI connected to a touch screen display for online monitoring. The procedure implemented for the determination of the detector efficiency is reported, along with the results of the measurements of the cosmic ray rate as a function of the altitude and the zenith angle, performed in the laboratory and in different locations outdoors.\\
}

\keywords{Particle detectors, Scintillators, scintillation and light emission processes (solid, gas and liquid scintillators), Solid state detectors}

\begin{document}
%
%
%
\maketitle
\flushbottom

\section{Introduction}
\label{sec:intro}
Cosmic ray detectors represent a singular opportunity for the development and execution of outreach experiences for high school and undergraduate students. This is due to the possibility to perform simple but effective measurements of several basic properties of particles and of radiation-matter interactions, without the need for complicated setups and radiation sources. Various experiences of this kind can be found in literature, such as \cite{AstrO,ArduSiPM,berkeley_assembly,berkeley_use,cosmic_watch,CosmO,geiger_cosmic,quarknet_ele}, where the setups generally focus on the stability, low price and ease of operation of the system.
The INSULAB portable cosmic ray detector was designed around the idea of a fully portable setup, from the detector itself to the electronics and the PC data acquisition.
As such, it has been assembled within a suitcase of light weight and compact dimensions, making it particularly suitable for outdoors experiences. The detector consists of three plastic scintillator modules coupled to photomultipliers which operate in coincidence and are readout by a custom compact electronics chain which allows the measurement of the arrival time and the time over threshold ($TOT$) of the signal generated by each scintillator module.\\
Compared to detectors that only operate in count mode, this setup offers the possibility of measuring and recording $TOT$ values that provide information on the deposited energy \cite{infoTOT}. Such values can be used to establish different offline cuts on the data, selecting the energy and direction of a given event.\\
Section \ref{sec:setup} offers a detailed description of the INSULAB cosmic ray detector, while section \ref{sec:char} outlines its characterization procedure. Finally, section \ref{sec:meas} reports the measurements performed with the detector itself.\\

\section{The portable detector}
\label{sec:setup}
\begin{figure}[htbp]
\centering 
\includegraphics[width=.48\textwidth]{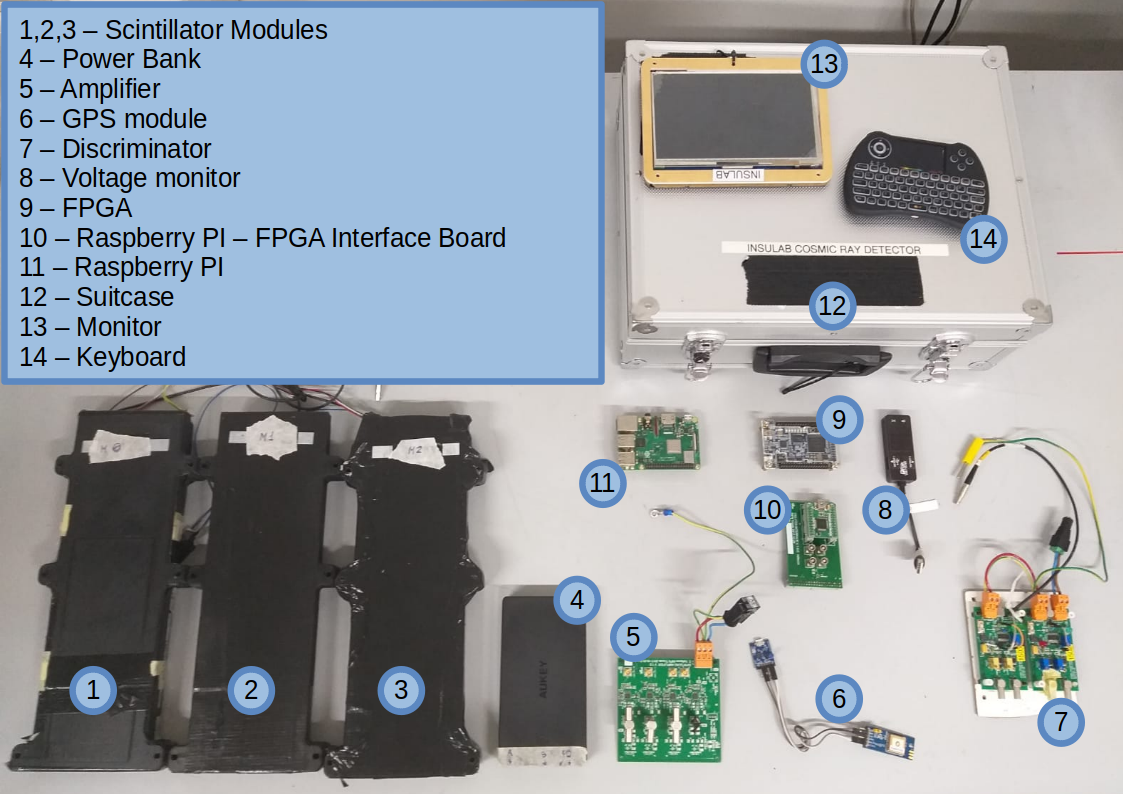}
\qquad
\includegraphics[width=.46\textwidth]{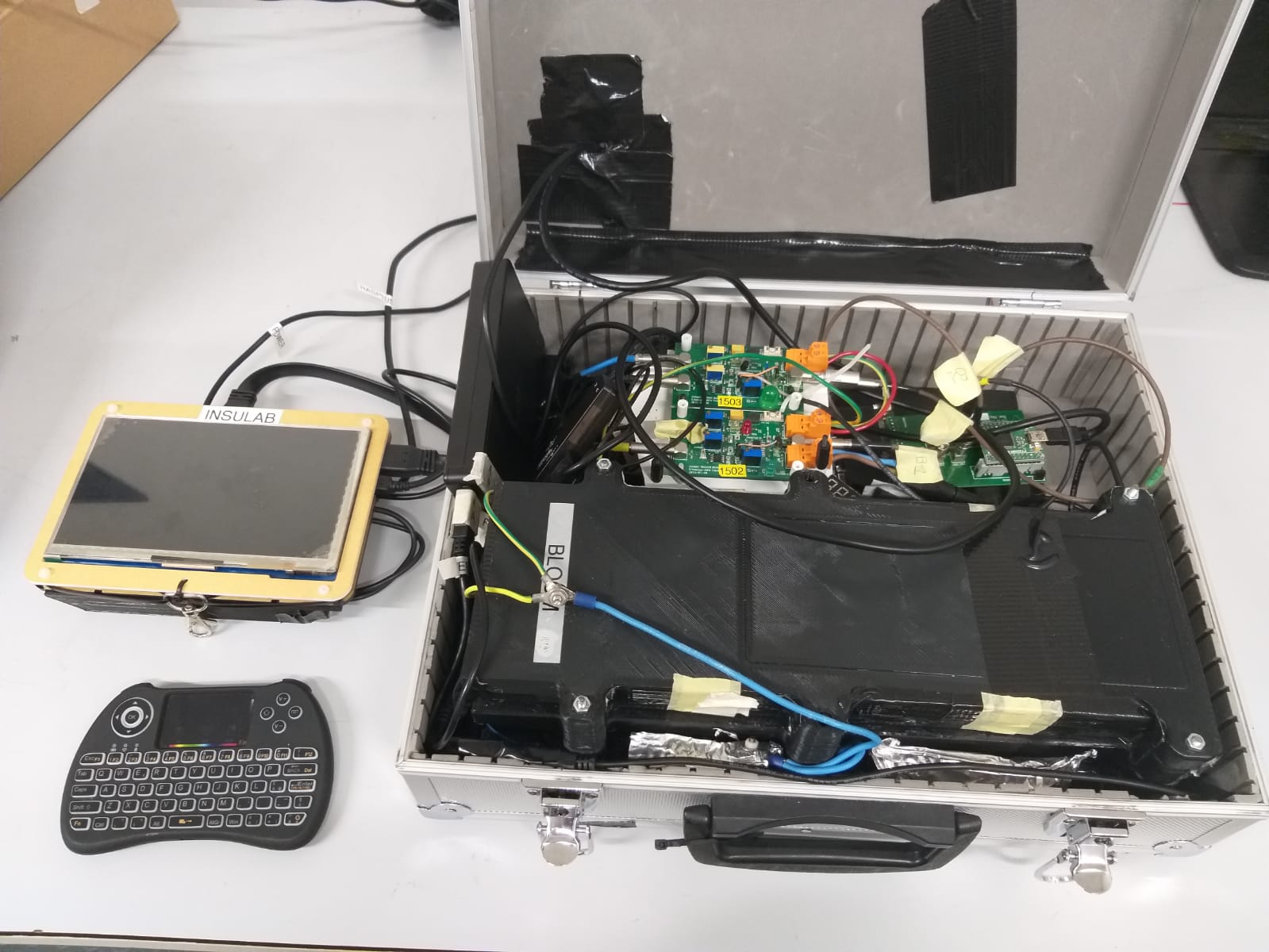}
\caption{\label{fig:IT} Left: individual components of the INSULAB portable detector. Right: view of the assembled detector, inside the suitcase.}
\end{figure}

\begin{figure}[htbp]
\centering 
\subfigure[]{\includegraphics[width=.53\textwidth]{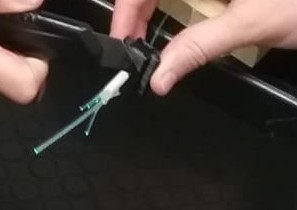}}
\subfigure[]{\includegraphics[width=.347\textwidth]{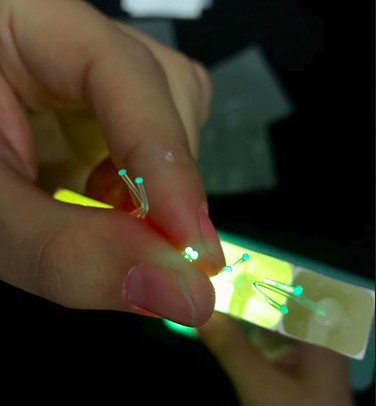}}
\subfigure[]{\includegraphics[width=.4\textwidth]{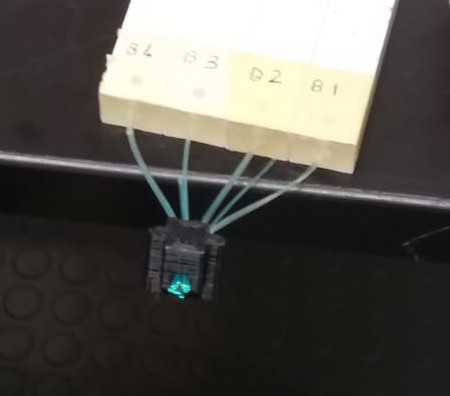}}
\caption{\label{fig:hole} (a) The fibers were inserted in the support, then cut with a hot blade; (b) after each cut and polishing the light yield was checked qualitatively by shining a light on one end; (c) the final assembly with the four bars.}
\end{figure}

\begin{figure}[htbp]
\centering 
\includegraphics[width=.75\textwidth]{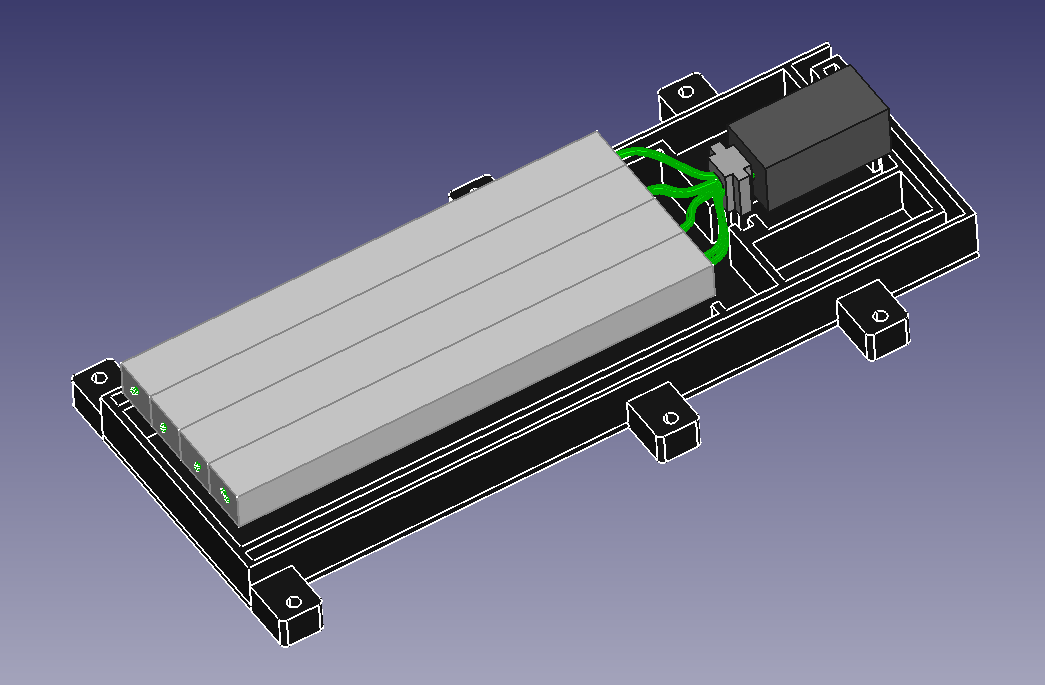}
\caption{\label{fig:cad_box} Drawing of an open PLA box with the scintillator bars, fibers, PMT and plastic support. }
\end{figure}

\begin{figure}[htbp]
\centering 
\includegraphics[width=0.75\textwidth]{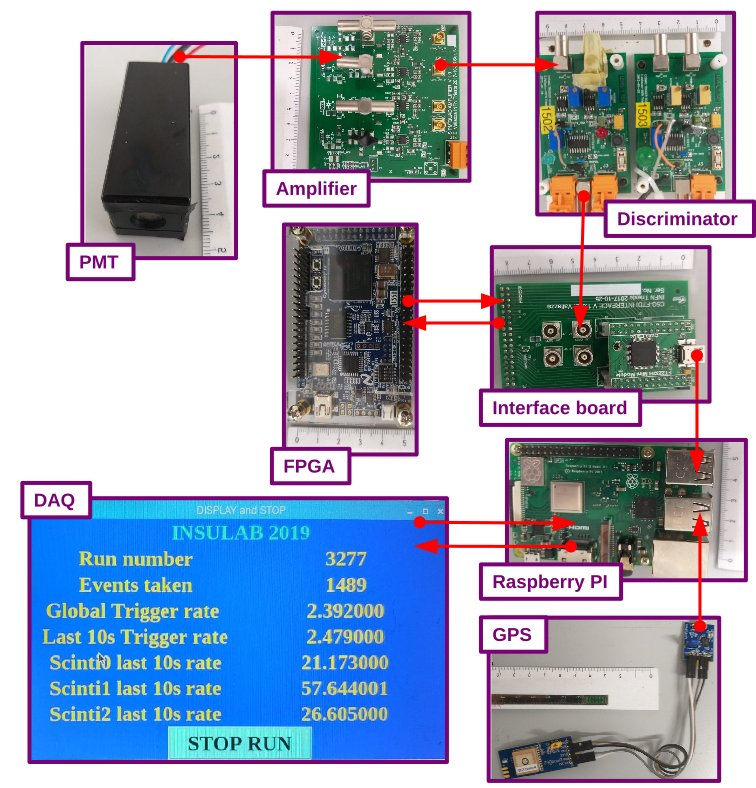}
\caption{\label{fig:electronics} Electronic chain components of the portable detector. See ruler (unit $=1$~cm) in the photos for the scale.}
\end{figure}

\begin{figure}[htbp]
    \centering
    \includegraphics[width=0.55\textwidth]{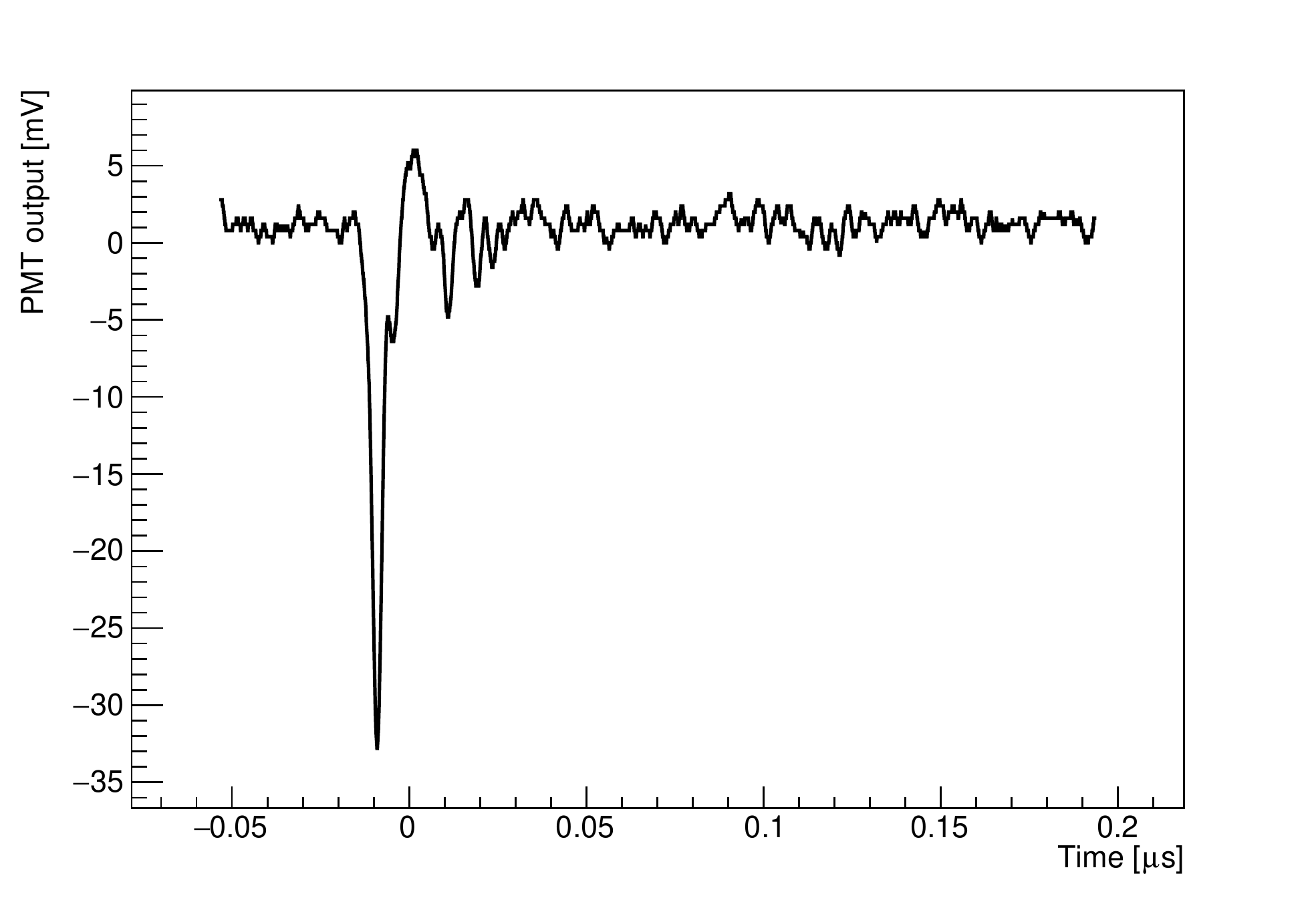}
    \caption{Example of a signal from one of the PMTs.}
    \label{fig:wfm_example}
\end{figure}

\begin{figure}[htbp]
    \centering
    \includegraphics[width=0.85\textwidth]{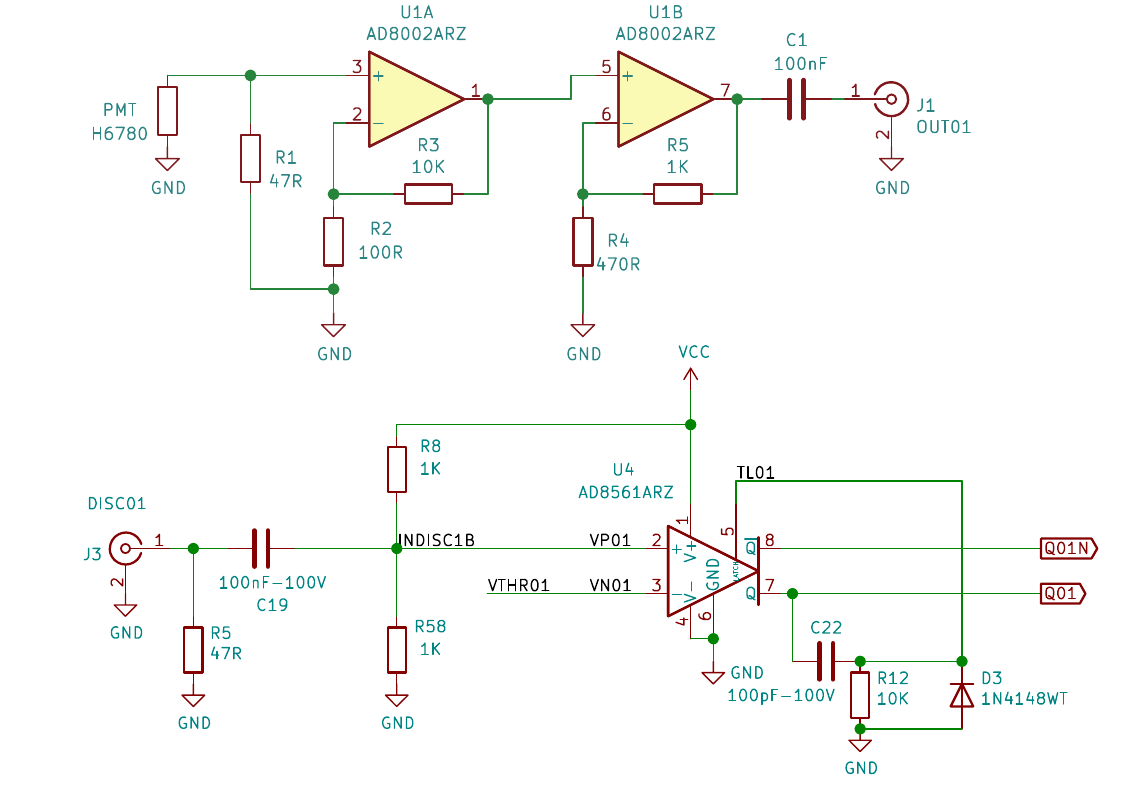}
    \caption{Schematic of the amplifier (top) and the discriminator (bottom). The shaping time of the amplifier is approximately 30~ns.}
    \label{fig:ele_schem}
\end{figure}

The INSULAB portable detector is shown in figure \ref{fig:IT}. The aluminum suitcase contains three 3D printed PolyLactic Acid (PLA) boxes, each containing four plastic scintillator bars and a single PhotoMultiplier Tube (PMT). The four scintillator bars in a box are coupled to the PMT through WaveLength Shifter (WLS) fibers. The electronics consists of an amplifier, a discriminator and a Field-Programmable Gate Array (FPGA). The PMTs operate with a 12~V power supply, obtained through a DC-DC converter from 5~V DC provided by a rechargeable power bank. The current and voltage for the whole electronics chain (nominal values: $V=$5~V, $I=$1.1~A) are displayed on a small screen inside the suitcase. A Raspberry PI 3 B+ \cite{Raspberrypi} connected to a small touch screen display \cite{display} and a portable mini-keyboard \cite{keyboard} controls the system and performs the data acquisition. In case of laboratory measurements, the detector can also be interfaced to any personal computer and powered with a common USB charger. A Digilent Pmod GPS receiver \cite{pmodgps} provides the GPS coordinates that are recorded, along with the current time, every few seconds during the data taking.
The plastic scintillators were provided by FNAL (Fermi National Accelerator Laboratory, USA) and are composed of a Dow Styron 663 (W) polystyrene base, doped with 1\% PPO $ + $ $0.03\%$ POPOP \cite{Beznosko2004}. Each bar has a cross-section of $1.5\times 1.9~$cm$^2$ and is 19.3~cm long and has a transmittance spectrum specified in \cite{pla2001low} and a peak of the emission spectrum at 410 nm \cite{asfandiyarov2016design}. A TiO$_2$ coating increases the light collection efficiency. They are readout via two 1~mm Kuraray WLS Y-11 (200M) \cite{Kuraray} fibers, glued in a hole inside the bar with the E-30 (PROCHIMA) epoxy resin \cite{PROCHIMA}. The fibers feature an absorption peak at 430~nm and an emission spectrum centered at 476~nm. Each $\sim 30$~cm long fiber runs along the full length of the bar and has an attenuation length larger than $3.5~$m. In each module, the 8 fibers are optically coupled to a $2.2 \times 2.2 \times 5$~cm$^3$ Hamamatsu H6780 \cite{Hamamatsu} PMT module. The Hamamatsu H6780 consists of a PMT coupled to a compact, high gain power supply circuit that allows operation of the PMT using a low voltage power source.\\
The detector was assembled following a series of steps; first of all the black PLA boxes were printed. In order to minimize leaks of ambient light, which might cause false triggers in the PMTs, the inside and outside surfaces of the boxes were covered with high coverage acrylic based paint (PRIMA, DUPLI-COLOR \cite{DUPLI-COLOR}). The fibers, already glued to the bars, were then cut with a hot blade in order to minimize micro fractures which could  compromise the light yield, and finally polished with sandpaper soaked in water. Some steps in the process are shown in figure \ref{fig:hole}. The fibers were then inserted into a plastic support, shown in figure \ref{fig:hole}c. This support has the function of holding the fibers in place in order to optimize the geometrical coupling with the PMT, while simultaneously covering the rest of the entrance window of the PMT, shielding it from potential light leaks coming from the front of the box. Optical grease was applied between the fibers and the PMT window to improve the optical coupling.\\
The scintillator bars, instrumented with the fibers and coupled to the PMTs, were finally positioned inside the boxes side by side, with the $1.5\times 19.3~$cm$^2$ faces of the bars in contact with each other. A drawing of the open box, showing the bars, the fibers, the PMT and the plastic support can be found in figure \ref{fig:cad_box}.
\\
The electronics chain is shown in figure \ref{fig:electronics}. The output signals from the PMTs (figure \ref{fig:wfm_example}) are small in amplitude (few~mV up to $\sim 200$~mV when terminated in 50~$\Omega$ resistor) and short (FWHM $\sim$10~ns), requiring amplification and shaping stages in the readout electronics chain.
The amplification is performed by a fast opamp (AD 8008, Analog Devices \cite{AD8008}) used in two non inverting stages with gain 100 and 3. The signal of each module is discriminated by a fast AD 8561 discriminator \cite{AD8561} with a threshold that can be manually adjusted through a potentiometer. A schematic of the amplifier and the discriminator is shown in figure \ref{fig:ele_schem}.
The discriminated signals are sent to the Altera Cyclone IV \cite{CycloneIV} FPGA housed in a DE0-Nano board from Terasic \cite{DE0}. 
The trigger signal $T$ for the system is generated by the majority of two out of three detectors.
When a trigger occurs, three 16-bit counters, that measure the time $\Delta t$ between two consecutive triggers, are stopped. The discriminated signals are sent to three 64-bit shift registers clocked at 200 MHz which record the time over threshold with a 5~ns sampling. The $TOT$ and $\Delta t$ data are finally transferred to the PC and recorded in ASCII files, alongside with the time of the operating system (with a $1~$s precision). The interface between the PC and the FPGA is a FT2232H Mini-Module \cite{FTDI} mounted on a custom board that connects its pins to the general purpose input/output pins on the Cyclone IV board. The custom board also features four connectors that deliver the discriminated signals to the FPGA. The data acquisition (DAQ) is performed by a program written in C, with a Tcl/Tk interface. The DAQ interface allows for easy start and stop of the acquisition and displays an estimate of the current trigger rate. The acquisition program can also measure the dead time $\tau$ of the data acquisition, which was determined to be 2.1~ms.\\
After closing the boxes and assembling the electronics chain, the whole system has been positioned inside the suitcase as shown in figure \ref{fig:IT}.

\section{Characterization of the detector}
\label{sec:char}
In this section we describe the measurements performed in order to determine the efficiency of the sctintillator modules and their response in terms of the time over threshold. 
\subsection{Measurement of the efficiency}
\label{sec:eff}
\begin{figure}[htbp]
\centering 
\includegraphics[width=.6\textwidth]{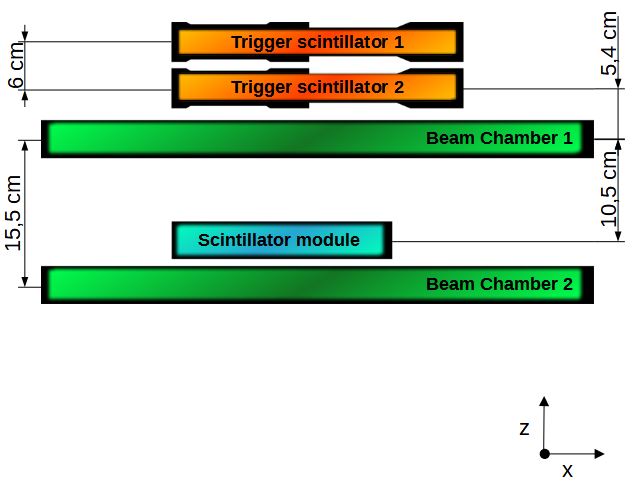}
\caption{\label{fig:effi} Schematic of the setup for the measurement of the efficiency.}
\end{figure}

\begin{figure}[htbp]
    \centering
    \includegraphics[width=0.95\textwidth]{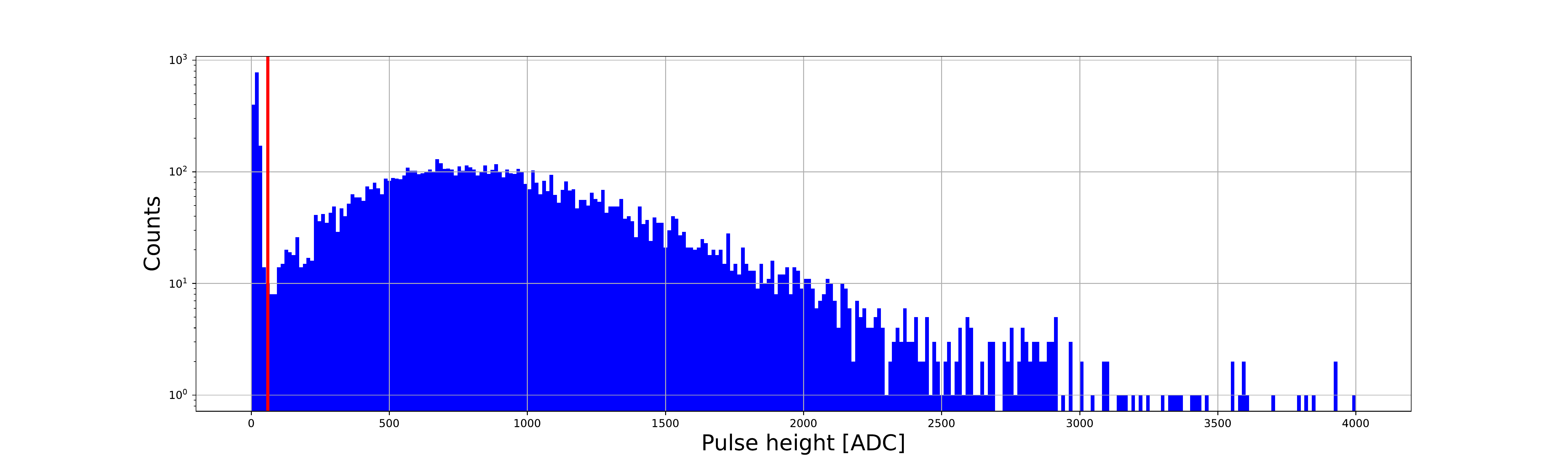}
    \caption{Example of a pulse height plot (1~ADC~$\simeq$~30.5~$\mu$V) , with the corresponding threshold to compute the efficiency (red line).}
    \label{fig:PHcut}
\end{figure}

\begin{figure}[htbp]
    \centering
    \includegraphics[width=1.0\textwidth]{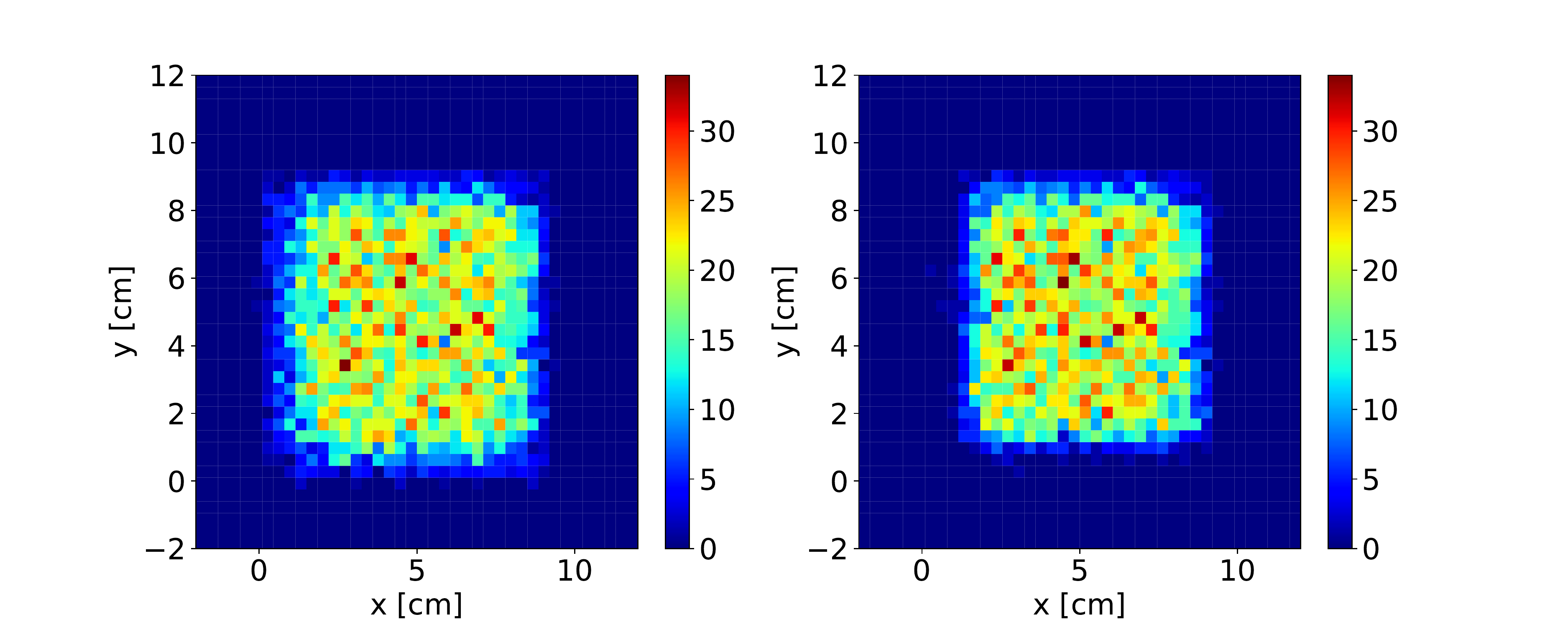}
    \caption{Left: 2D histogram of the ($x$, $y$) hit positions projected onto the plane of the scintillator module. Right: 2D map of the events detected by the scintillator module. The pixel size is $0.25\times 0.25$~cm$^2$.}
    \label{fig:eff_proc}
\end{figure}

\begin{figure}[htbp]
\centering
\includegraphics[width=0.95\textwidth]{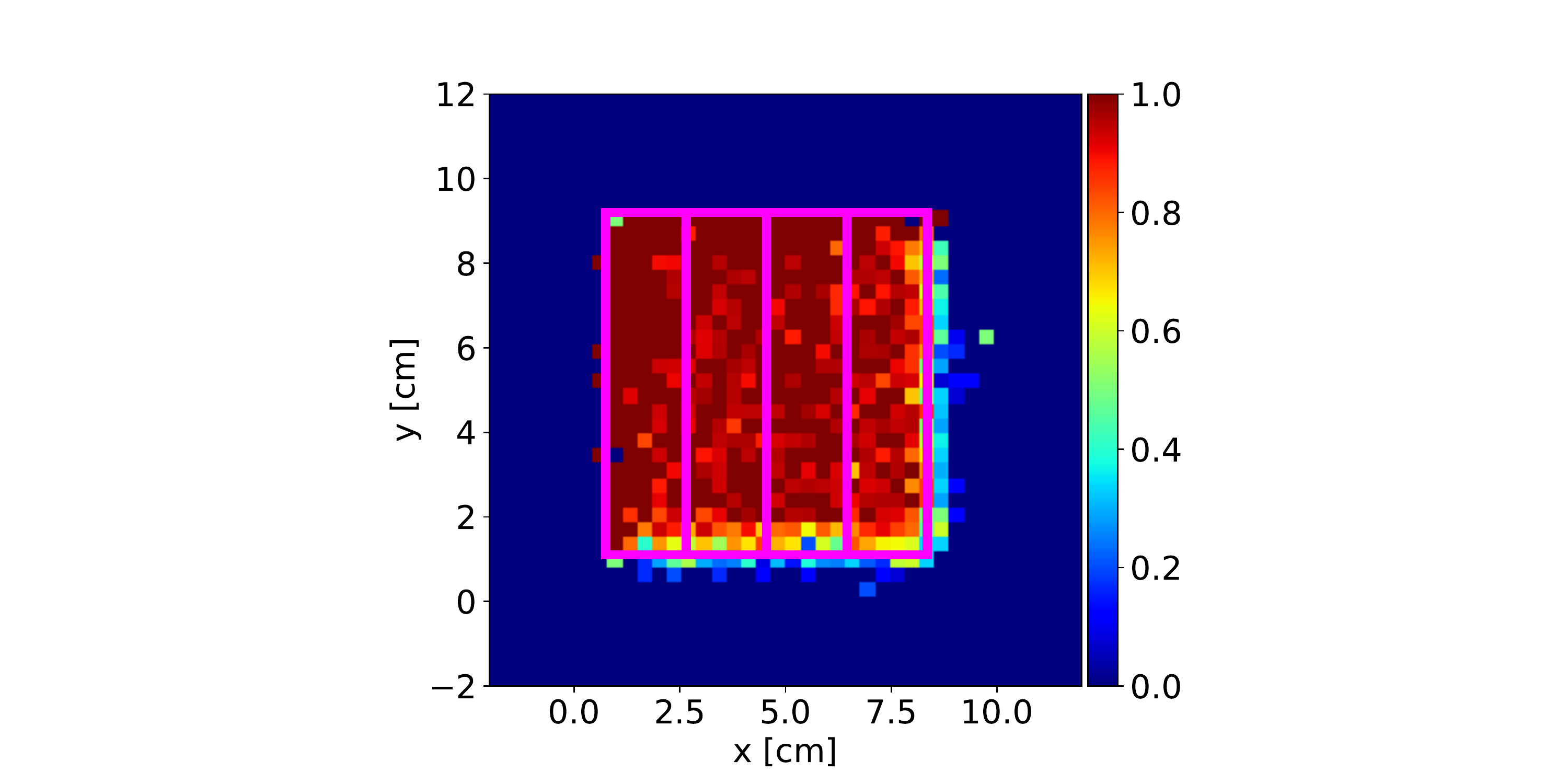}
\qquad
 \includegraphics[width=.65\textwidth]{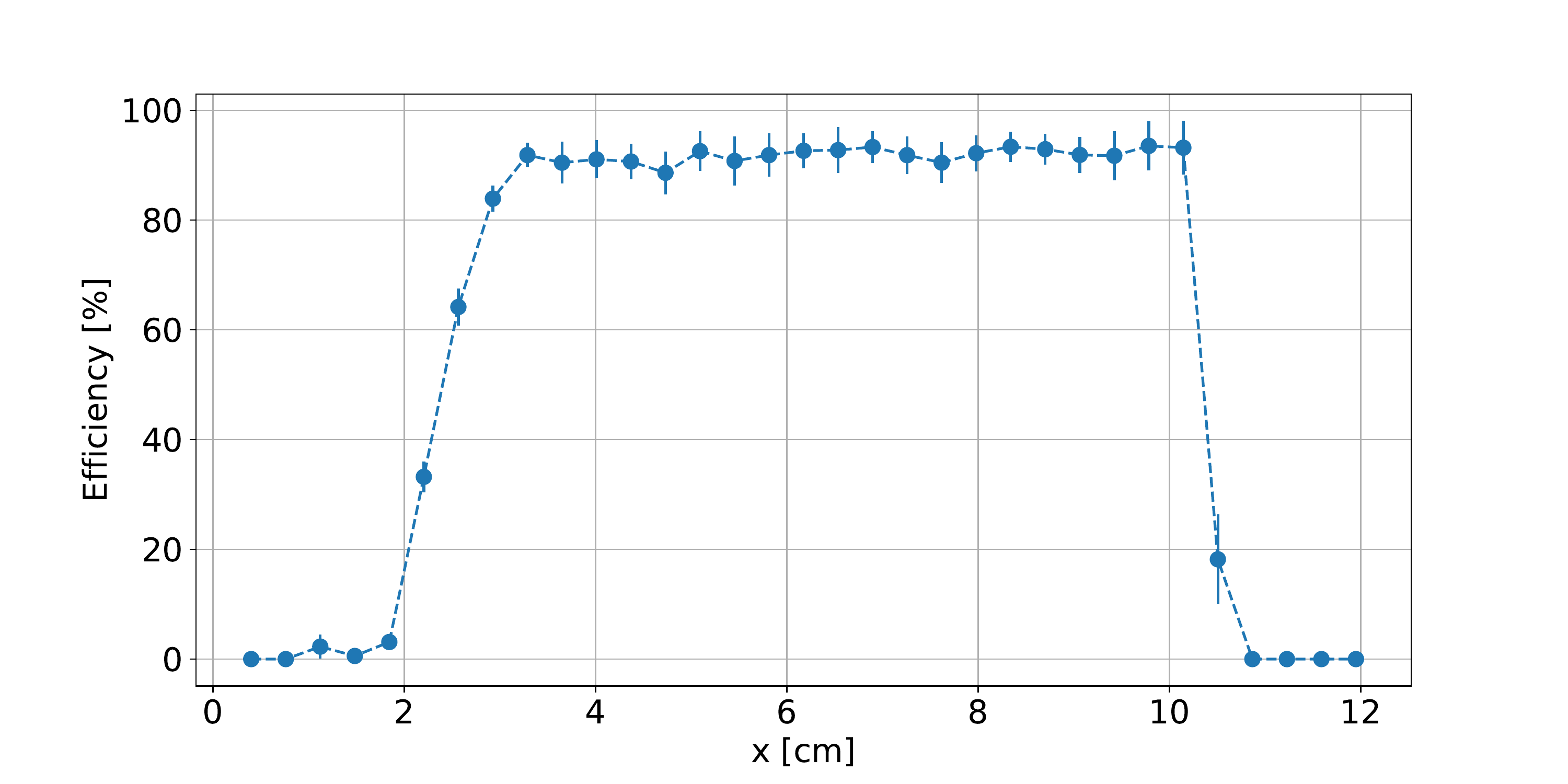}
\caption{\label{fig:eff_prof} Top: 2D efficiency map of a scintillator module. The pixel size is $0.25\times 0.25$~cm$^2$. The lines indicate the actual position of the scintillator bars. Bottom: efficiency profile along the $x$ coordinate. }
\end{figure}

The characterization of the detector in terms of the efficiency of each scintillator module has been performed in the laboratory using cosmic rays. The setup is shown in figure \ref{fig:effi}.
The coincidence of two $10\times 10$~cm$^2$ plastic scintillators provides the trigger. Two silicon Beam Chambers (BCs) are positioned below the scintillators. Each BC consists of two single side silicon microstrip detectors \cite{AGILE} in an $x$, $y$ configuration. The hit positions on the microstrips are recorded and used to reconstruct the track of the incoming cosmic ray. The features of the BC silicon detectors are reported in table \ref{tab:strips}.\\
To measure the efficiency, each scintillator module is positioned between the two BCs and its signal, shown in figure \ref{fig:wfm_example}, is sampled with a CAEN DT5730 14 bit 500~MS/s Digitizer module \cite{CAENdigi}.\\
The data acquisition software processes the digitized waveform, selecting the maximum value and recording it. By putting a threshold on the pulse height (figure \ref{fig:PHcut}), one can determine the number of successfully detected events.\\
The tracks are projected onto the horizontal plane corresponding to the scintillator module, obtaining a set of ($x$, $y$) coordinates, which are stored in a bidimensional histogram (figure \ref{fig:eff_proc}, left). Events with multiple hits on the BCs are discarded. After applying the pulse height cut shown in figure \ref{fig:PHcut}, the 2D map (figure \ref{fig:eff_proc}, right) of the events detected by the scintillator is produced. The ratio between such map and the initial 2D histogram provides a 2D map of the efficiency of the scintillator module.
\begin{table}[htbp]
\centering
\begin{tabular}{|c|c|}
\hline
Numbers of readout strips & 384\\
\hline
Thickness & 410$~\mu$m \\
\hline
Dimensions& $9.5\times 9.5~$cm$^2$\\
\hline
Strip physical pitch & 121$~\mu$m \\
\hline
Strip readout pitch & 242$~\mu$m\\
\hline
\end{tabular}
\caption{\label{tab:strips} BC silicon detectors characteristics \cite{AGILE}.}
\end{table}
The efficiency map of one of the three modules, along with its projection on the $x$ axis, is shown in figure \ref{fig:eff_prof}. The value of the efficiency of each module, obtained by averaging the maps, is reported in table \ref{tab:efftab}. The thresholds of the discriminator were set to the same values used for the pulse height cut; if the operating temperature range is within the specification given by the manufacturer for the PMT (5-45$^\circ$~C), we expect the efficiency in the suitcase to be similar to the one measured in the lab.
\begin{table}[htbp]
\centering
\begin{tabular}{|cc|}
\hline
Module & Efficiency $\%$\\
\hline
Module 0 & 90.3 $\pm$ 1.4\\

Module 1 & 84.2 $\pm$ 1.9\\

Module 2 & 96.4 $\pm$ 1.1\\

\hline
\end{tabular}
\caption{\label{tab:efftab} Efficiency values of the three modules. Modules 0 and 1 are slightly less efficient than module 2, probably due to some of the fibers being damaged during the assembly. }
\end{table}

\subsection{Measurement of the time over threshold response}
\label{sec:TOT}
\begin{figure}[ht]
\centering
\includegraphics[width=0.65\textwidth]{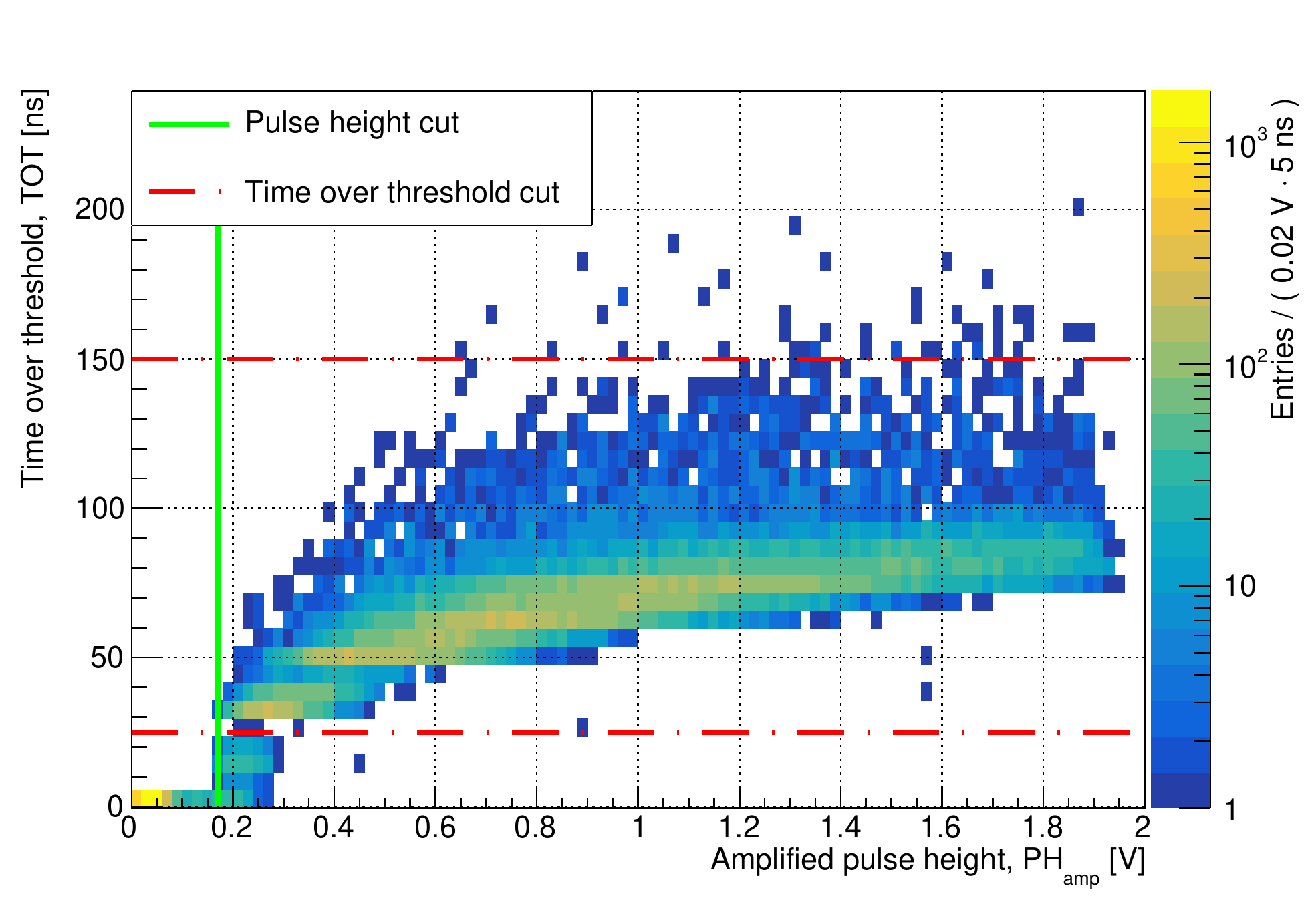}
\caption{\label{fig:TOTvsPH} Time over threshold plotted as a function of the amplified pulse height for module 0, showing the pulse height and $TOT$ cuts.}
\end{figure}

The INSULAB portable detector provides the $TOT$ of each event. This quantity can be used to perform offline quality cuts, as described in section \ref{sec:meas}, in order to select the events corresponding to the crossing of a particle. In order to establish such cuts, the relationship between the pulse height and the $TOT$ quantities was investigated with a cosmic ray run. The scintillator modules were positioned on top of each other, in the same configuration adopted for the suitcase. The amplified waveforms of all the three modules, along with the discriminated waveforms of module 0, were digitized by the CAEN DT5730 digitizer. As the measurement focused on module 0, the trigger for the acquisition was the logic OR between its discriminated waveform and the amplified waveforms of the two other modules, discriminated with a high ($\sim$1.8~V) threshold. This choice was made in order to guarantee high statistics for module 0, while simultaneously acquiring a sample of events in which it was not triggered (namely, those in which only the other two modules satisfied the trigger conditions).\\
The amplified pulse height of every module, along with the $TOT$ for module 0, was computed for each event. The population of events detected by at least two of the three modules, approximately mimicking the trigger logic of the suitcase system, was selected via a cut on the amplified pulse heights. For such events, the $TOT$ of module 0 is shown in figure \ref{fig:TOTvsPH} as a function of the amplified pulse height. A selection on the $TOT$ was established. The lower bound was chosen in order to exclude events with a low pulse height, selecting only those in which module 0 was actually triggered. The purpose of the upper bound is to reject electronic noise, which might result in a long discriminated signal that does not necessarily correspond to a real trigger. While such events are not abundant in the data presented in this section, some external sources can cause such an effect. Among the 30960 events detected by at least two of the modules, 26733 had a pulse height above the threshold set on channel 0, while 26187 had a $TOT$ value within the established cut, all of which are also included in the pulse height selection. As such, we conclude that the cut based on the $TOT$ values adequately selects the population of actual triggers of a certain channel, with a loss of efficiency of no more than a few \%. The measurement was repeated for modules 1 and 2, yielding similar results.
\section{Cosmic ray measurements}
\label{sec:meas}
\begin{figure}[ht]
    \centering
    \subfigure{\includegraphics[width=0.75\textwidth]{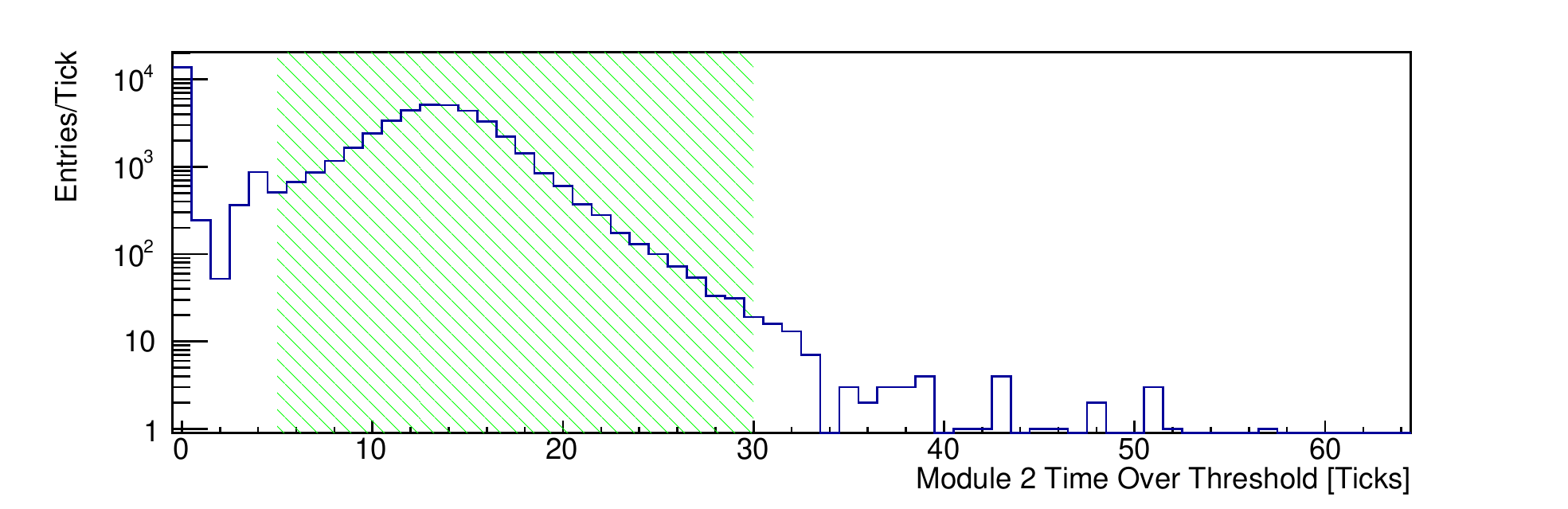}}
    \subfigure{\includegraphics[width=0.75\textwidth]{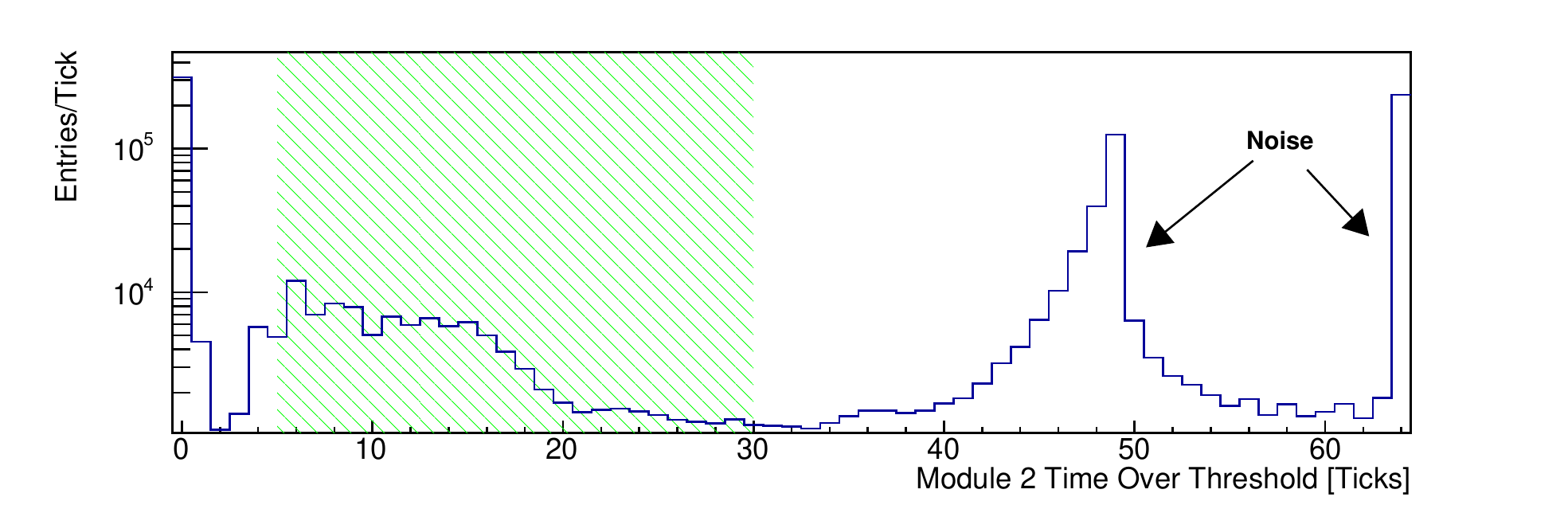}}
    \caption{Top: example of a typical $TOT$ spectrum, with one tick corresponding to 5~ns. The highlighted area indicates the selection applied for the triple coincidence requirement. Bottom: example of a $TOT$ spectrum heavily affected by external noise. The upper bound of the selection is chosen in order to exclude such noise.}
    \label{fig:TOTcut}
\end{figure}

\begin{figure}[htbp]
    \centering
    \subfigure{\includegraphics[width=0.48\textwidth]{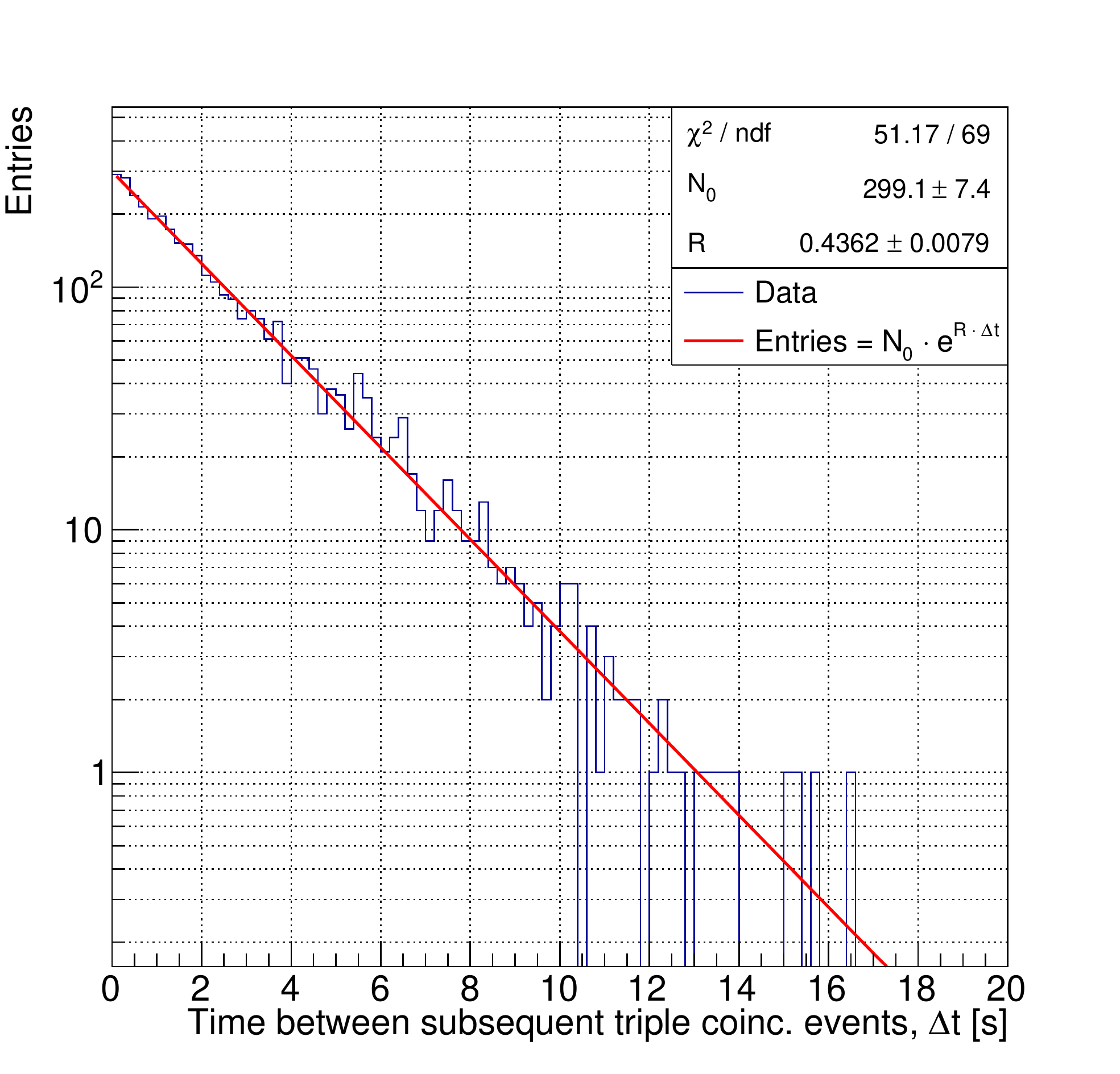}}
    \subfigure{\includegraphics[width=0.48\textwidth]{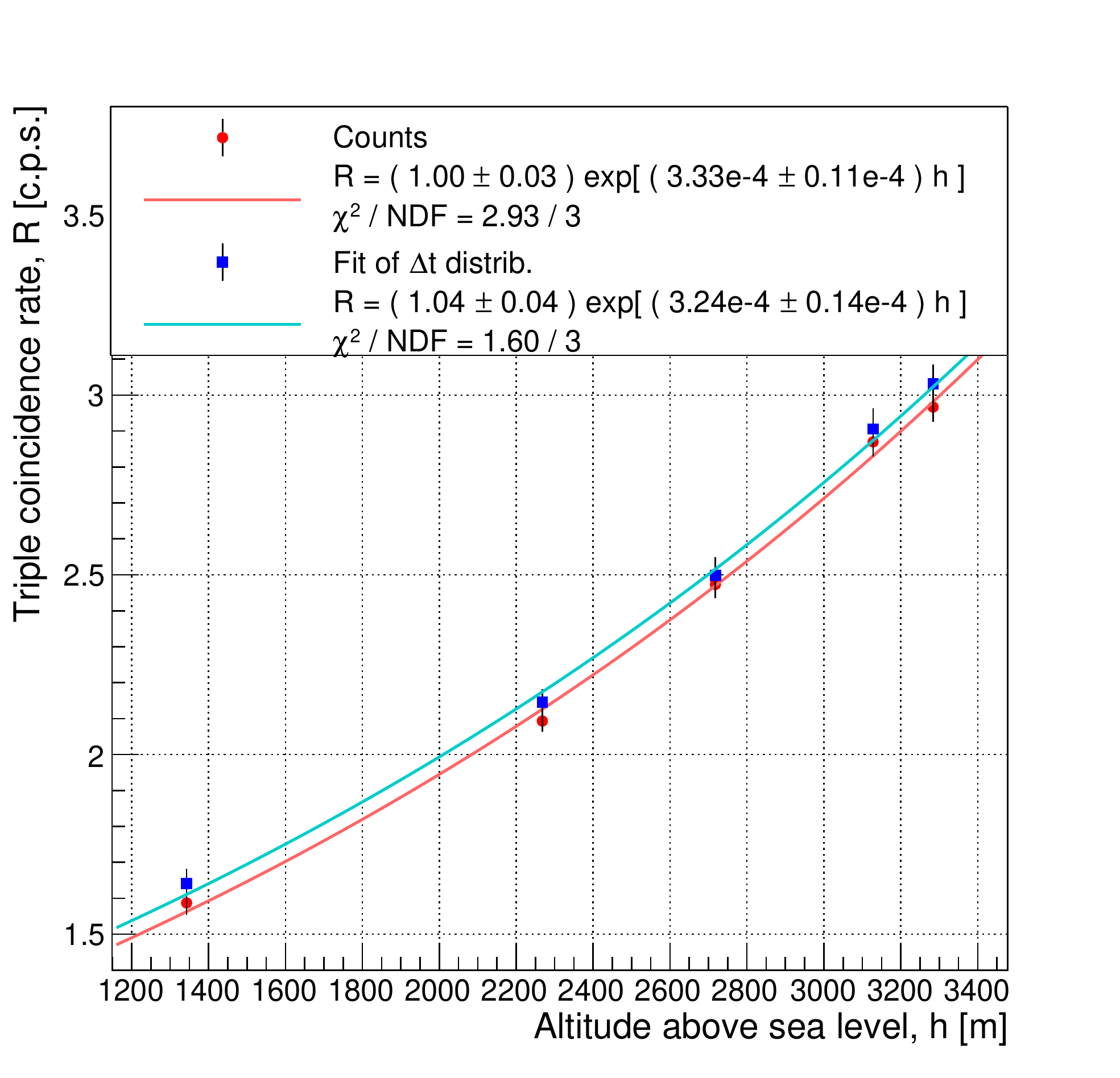}}
    \caption{Left: distribution of the time intervals $\Delta t$ between subsequent triple coincidence events. The exponential fit yields a rate of $\left( 0.436\pm 0.008\right)~$Hz, compatible within less than 1$\sigma$ with the $\left( 0.434 \pm 0.007 \right)~$Hz value obtained from the analysis of the triple coincidence counts. Right: count rate as a function of altitude for the Stelvio pass run; comparison between the two methods. Both data sets are well fit by an exponential function, with all the data points and fit parameters being compatible within $1\sigma$. }
    \label{fig:exp_rate}
\end{figure}

\begin{figure}[htbp]
    \centering
    \includegraphics[width=1.0\textwidth,keepaspectratio]{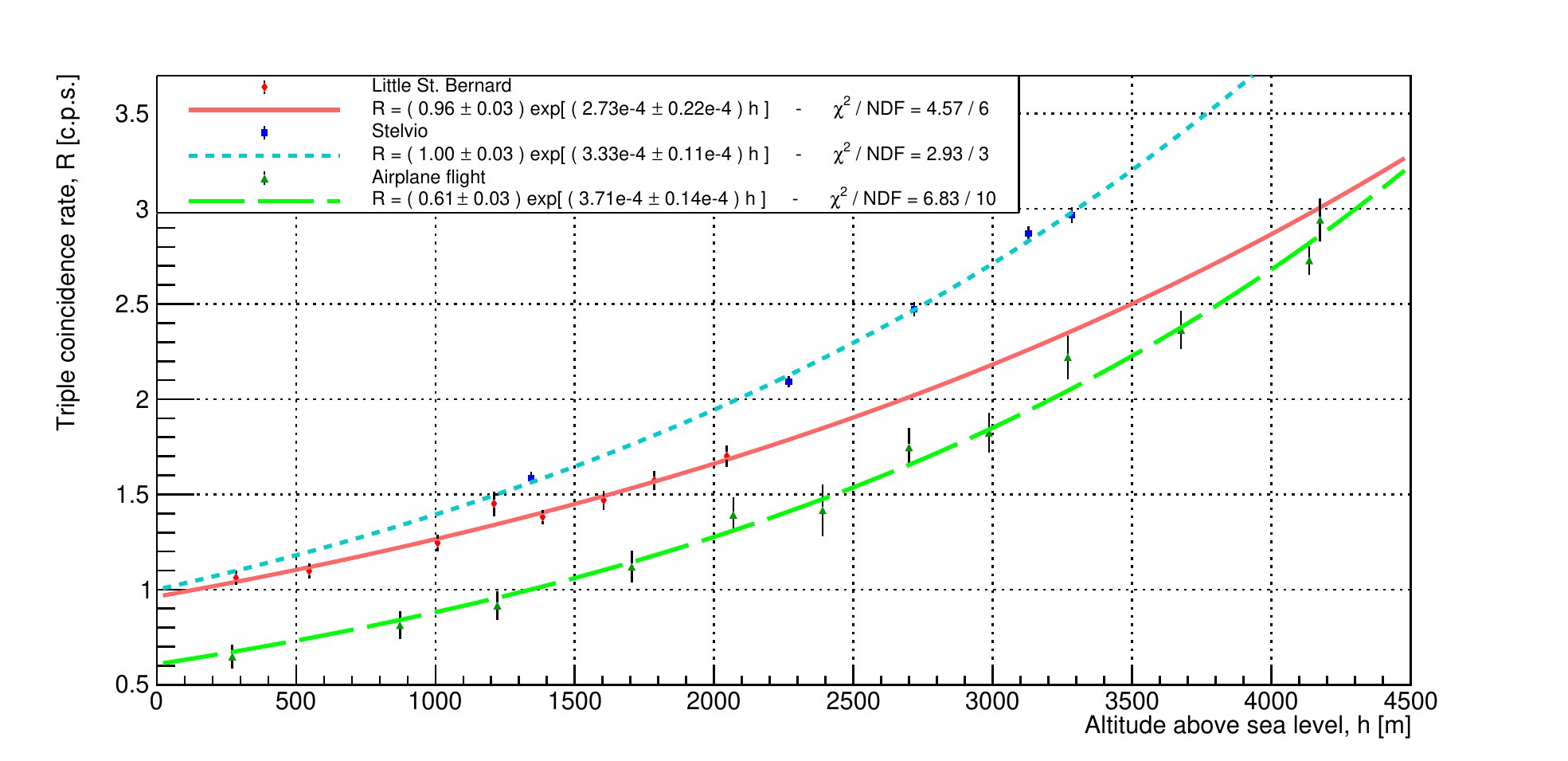}
    \caption{Cosmic ray rate as a function of altitude. The data sets were taken in different conditions (inside or outside of a vehicle, different configurations of the electronics chain), so no meaningful comparison can be made between different data sets. Each set, however, follows the expected exponential growth.}
    \label{fig:flux_vs_alt}
\end{figure}

\begin{figure}[htbp]
    \centering
    \includegraphics[width=0.75\textwidth,keepaspectratio]{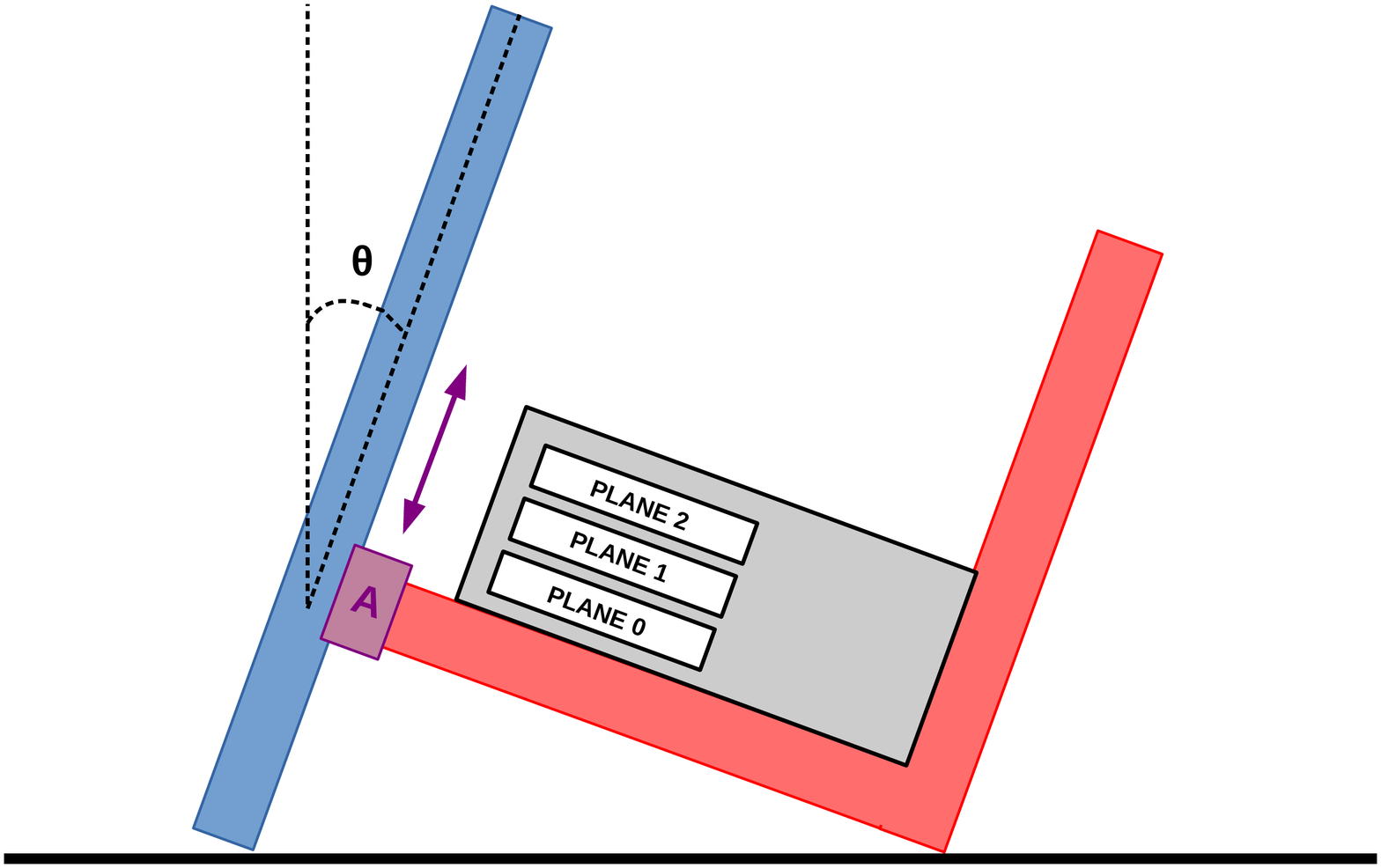}
    \caption{Schematic of the setup for the measurement of the cosmic ray rate as a function of the zenith angle $\theta$. The fixing point $A$ between the red L-shaped support and the long blue bar can be changed, thus varying $\theta$. }
    \label{fig:angle_setup}
\end{figure}

\begin{figure}[htbp]
    \centering
    \includegraphics[width=1.0\textwidth,keepaspectratio]{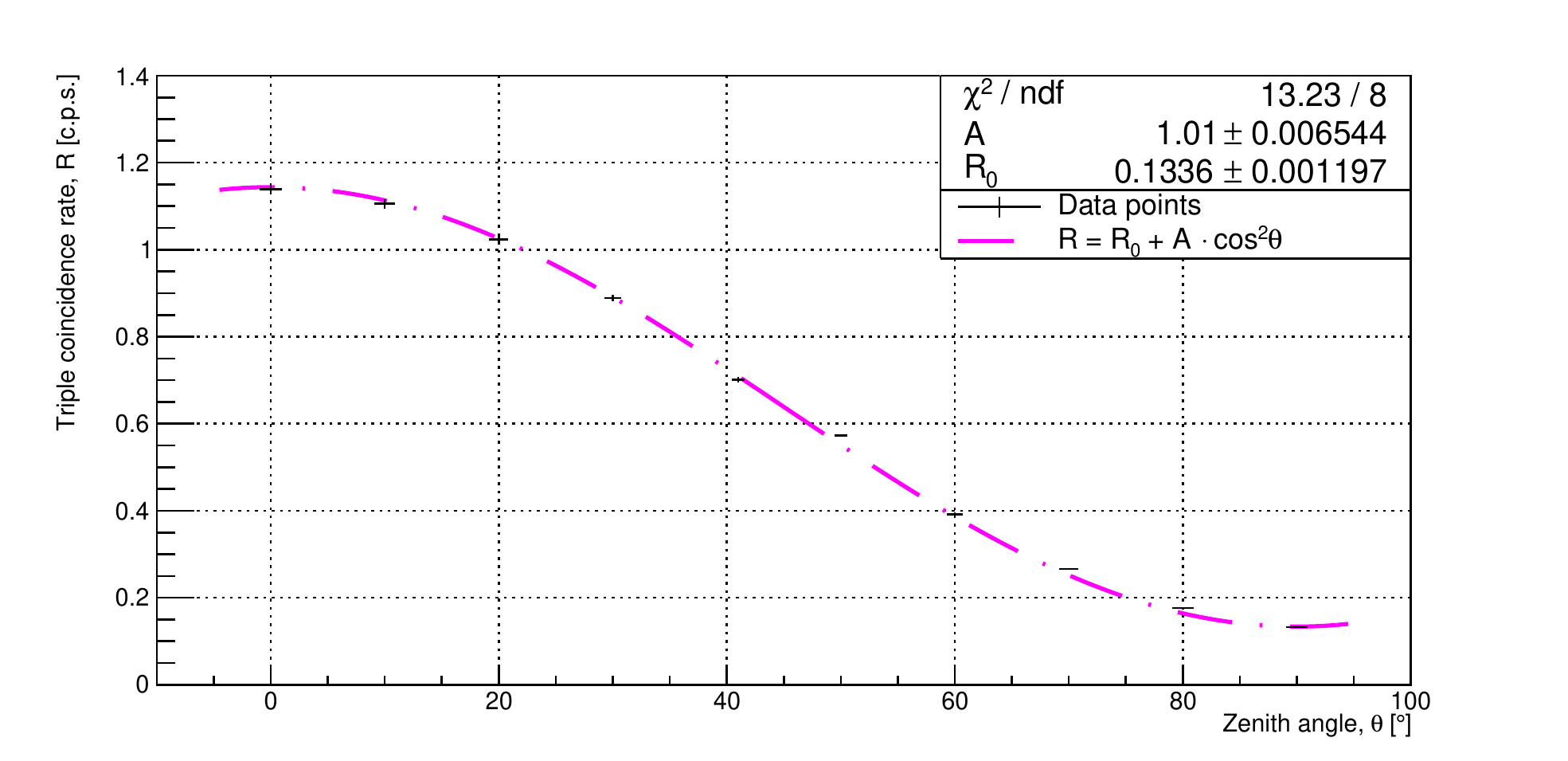}
    \caption{Cosmic ray rate as a function of the zenith angle $\theta$. The uncertainty on $\theta$ is estimated considering the limited precision of the positioning of the apparatus in figure \ref{fig:angle_setup}.}
    \label{fig:flux_vs_ang}
\end{figure}

\begin{figure}[htbp]
    \centering
    \includegraphics[width=1.0\textwidth,keepaspectratio]{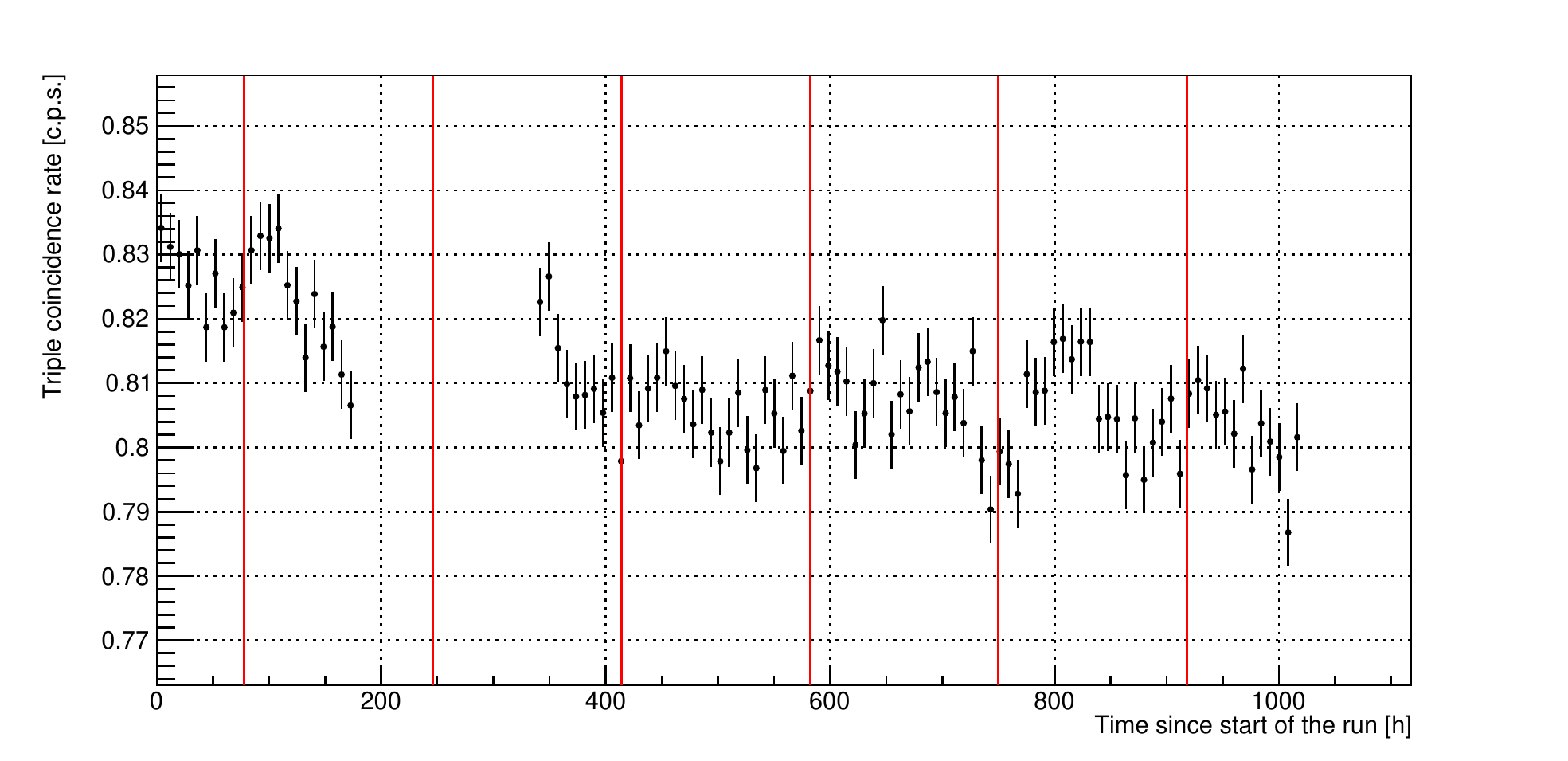}
    \caption{Cosmic ray rate as a function of time. The data were taken in fixed detector conditions for 42 days, with a $\sim$6 days long break after the first week. The red vertical lines correspond to the start of each new week. Small - but statistically significant - variations in the rate can be observed through the course of the data taking. }
    \label{fig:r_long_run}
\end{figure}

Two well-known behaviours of cosmic rays have been successfully observed with the portable detector: the increase of the cosmic ray flux with altitude and its $\cos ^2\theta$ modulation \cite{PDG} as a function of the orientation $\theta$ with respect to the vertical axis. Such measurements can be easily repeated and constitute a solid basis for the development of a future didactic experience for high school students. In 2019 and 2020 the INSULAB cosmic ray detector has been taken on several trips, acquiring cosmic ray data in various locations, reported in table \ref{tab:alt_runs}, that cover a wide range of altitudes. In each run, background events can be generated by light leaks and dark counts of the PMTs. Moreover, electronic noise from external sources can generate false counts. These events are excluded by selecting the ``triple coincidence'' events - that is, the events in which all the three scintillator modules detected a particle. The selection is performed by requiring that the $TOT$ of each module lies within the interval of values selected in section \ref{sec:TOT} (see figure \ref{fig:TOTcut}).
The number of triple coincidence events is then divided by the total time $T_{run}$ of the run, obtaining the count rate $r$. Lastly, the real rate $R$ is computed by applying a correction to account for the dead time $\tau$ of the acquisition:
\begin{align*}
    R=r\left( 1-N\frac{\tau}{T_{run}} \right)^{-1}
\end{align*}
where $N$ is the total number of triggers in the run. The dead time correction is relatively small (between $\sim$0.2\% and $\sim$1.5\%), assuming standard run conditions of few triggers per second and no significant contribution from external noise.
An alternative method to estimate the cosmic ray count rate involves the study of the distribution of the time interval, $\Delta t$, between subsequent events. The value recorded by the data acquisition does not include the dead time needed to save the data and reset the system after the previous trigger. Therefore, the real time interval between triggered events is the sum of the value given by the DAQ and the dead time mentioned above. The distribution of the $\Delta t$ values can be fit by an exponential function (figure \ref{fig:exp_rate}, left) due to the poissonian nature of the process where the decay constant yields an estimate of the rate. An example can be seen in figure \ref{fig:exp_rate} (right), where a comparison between the results of both the analyses is presented: the resulting cosmic ray rates are compatible.
\begin{table}[htbp]
    \centering
    \begin{tabular}{|l|l|l|}
        \hline
        Destination & Altitude above sea level $\left[\mbox{m}\right]$ & Collected events  \\
        \hline
        Little St. Bernard Pass, France \& Italy & 133 to 2187 & 73818\\
        Stelvio pass, Italy & 1343 to 3284 & 54592 \\
        Airplane flight & 269 to 4178 & 52798\\
        \hline
    \end{tabular}
    \caption{Cosmic ray data acquisition in different trips. }
    \label{tab:alt_runs}
\end{table}
The cosmic ray rate as a function of the altitude above the sea level, as measured in the runs reported in table \ref{tab:alt_runs}, is shown in figure \ref{fig:flux_vs_alt}. A precise prediction of the cosmic muon flux as a function of altitude requires complex calculations that model the primary cosmic ray spectrum, the production of muons starting from pions, the behaviour of the atmosphere in terms of local density and its interaction with the muons. At relatively low altitudes ($<$10~km) the flux, as estimated in works such as \cite{Lipari}, can be reasonably approximated by an exponential growth with altitude. As seen in figure \ref{fig:flux_vs_alt}, the data are well fit by an exponential function.
The second measurement involved positioning the detector at different orientations, through a simple setup shown in figure \ref{fig:angle_setup}, varying the zenith angle $\theta$ between 0 and 90$^\circ$.
For each value of $\theta$, the count rate was determined with the same methods as in the altitude runs. The data points, shown in figure \ref{fig:flux_vs_ang}, were fit with the function
\begin{align*}
    R=R_0+A\cos ^2\theta
\end{align*}
%
where the $R_0$ term accounts for the large angular acceptance resulting from the wide surface of each scintillator module, that results in a non-zero rate, even at a 90$^\circ$ angle.
As shown by the fairly large $\chi^2/NDF$ ratio, the function does not fit the data points perfectly. This is probably due to the fact that the data taking was performed across several days, in order to gather enough statistics for the large angle (and consequently low rate) points. As the cosmic rays flux at ground level is influenced by the local atmospheric conditions, the data sets taken across several days may not be fully comparable. The effect can be seen in figure \ref{fig:r_long_run}, where the evolution of the cosmic ray rate in fixed detector conditions is shown through a period of $\sim$42 days. A possible future improvement to the system might include the ability to correlate the rate to the local atmospheric pressure, as was done in \cite{CosmO_pressure}.

\section{Conclusions}
In this paper we have presented a description of the INSULAB portable cosmic ray detector, along with the data for its characterization and the first tests performed with it. This detector was first conceived in the context of the ``Laboratorio di Fisica IV'' course, during which the third year bachelor degree students at Universit\`a degli Studi dell'Insubria perform nuclear and subnuclear physics experiments. It was later fully assembled and characterized as part of a bachelor degree thesis at Universit\`a degli Studi dell'Insubria \cite{TesiAle}. The detector proved to be effective in measuring some well-known behaviours of cosmic rays, such as the dependence of the rate on the altitude and on the zenith angle.  These measurements can be implemented in an educational experience for high school students, thanks to the portability of the system - which fits in a suitcase - and the simple Tcl/Tk interface, which allows to start and stop the acquisition with the press of a single button.
With the (as of this time) planned reopening of Italian high schools in september/october 2021, we plan to organize activities that involve the detector in several schools in northern Italy.

\acknowledgments
We would like to thank Alessandro Kosoveu (INFN-Ts) for the original design of the PLA boxes and Luigi Bomben for modifying it into the final version and 3D printing the boxes. Our thanks also go to the Laboratorio di Fisica IV class of 2018 (Camilla Bianciardi, Giulia Conenna, Federica Galli, Aurora Luppi, Sofia Maggioni and Margherita Marazzi), Giovanni Ballerini, Sofia Mangiacavalli and Mattia Ciliberti, who all participated in part of the assembly and data taking.

\bibliographystyle{JHEP}
\bibliography{bibliography} 
\end{document}